\documentclass{llncs}

\usepackage{llncsdoc}
\usepackage{graphicx}

\newcommand{\keywords}[1]{\par\addvspace\baselineskip
\noindent\keywordname\enspace\ignorespaces#1}

\begin {document}
\title{A Dynamic Phase Selection Strategy for Satisfiability Solvers}
\titlerunning{Phase Selection for SAT Solvers}

\author{Jingchao Chen}
\institute{School of Informatics, Donghua University \\
2999 North Renmin Road, Songjiang District, Shanghai 201620, P. R.
China \email{chen-jc@dhu.edu.cn}}

\maketitle
\begin{abstract}

The phase selection is an important of a SAT Solver based on
conflict-driven DPLL. This paper presents a new phase selection
strategy, in which the weight of each literal is defined as the sum
of its implied-literals static weights. The implied literals of each
literal is computed dynamically during the search. Therefore, it is
call a dynamic phase selection strategy. In general, computing
dynamically a weight is time-consuming. Hence, so far no SAT solver
applies successfully a dynamic phase selection. Since the implied
literal of our strategy conforms to that of the search process, the
usual two watched-literals scheme can be applied here. Thus, the
cost of our dynamic phase selection is very low. To improve Glucose
2.0 which won a Gold Medal for application category at SAT 2011
competition, we build five phase selection schemes using the dynamic
phase selection policy. On application instances of SAT 2011,
Glucose improved by the dynamic phase selection is significantly
better than the original Glucose. We conduct also experiments on
Lingeling, using the dynamic phase selection policy, and build two
phase selection schemes. Experimental results show that the improved
Lingeling is better than the original Lingeling.

\end{abstract}

\section{Introduction}
As a classic NP-complete problem, Satisfiability (SAT) has been
studied for for a long time. Numerous state-of-the-art solvers have
been developed in order to solve some problems in the fields such as
computer aided design, data diagnosis, EDA, logic reasoning,
cryptanalysis, planning, equivalence checking, model checking, test
pattern generation etc. However, now large real-world SAT problems
remain unsolvable yet.

 In general, SAT solvers are classified into conflict-driven, look-ahead
and random search. Solvers on application instances are almost all
conflict-driven DPLL-type solvers. This paper focuses on this type
of solvers. A conflict-driven DPLL-type solver consists of variable
selection, phase selection, BCP (Boolean Constraint Propagation),
conflict analysis, clause learning and its database maintenance. The
optimization of each component is useful for improving the
performance of solvers. So far, Numerous optimizing strategies has
been proposed. For example, for variable selection, the
corresponding optimizing strategy is VSIDS (Variable State
Independent Decaying Sum) scheme \cite{CDCL:16}. To accelerate BCP,
two watched-literals scheme was proposed. With respect to conflict
analysis, a large amount of optimizing work has been done. For
example, firstUIP (unique implication points), conflict clause
minimization, on-the-fly self-subsuming resolution \cite{Onfly:17},
learned clause minimization \cite{Min:18} etc are used to optimize
conflict analysis. To maintain effectively clause learning database,
in 2009, Audemard et al. introduced a Glucose-style reduction
strategy \cite{glue:19} to remove less important learned clauses. In
2011, they presented further a freezing and reactivating policy
\cite{Audemard:2011} to restore the most promising learnt clauses
rather than to re-compute them. Due to this new technique, Glucose
2.0  won a Gold Medal for application category at SAT 2011
competition.

  Unlike the other components of conflict-driven SAT solvers, the literature
on the phase selection of variables is rare. To our best knowledge,
up to now, only two phase selection strategies are widely used in
conflict-driven SAT solvers. One is the phase selection heuristic
used in RSAT (RSAT heuristic for short) \cite{RsatHeuris:1}. The
other is Jeroslow-Wang heuristic \cite{JWheuris:2}. The basic idea
of the RSAT heuristic is to save the previous phase and assign the
decision variable to the same value when it is visited once again.
The basic idea of Jeroslow-Wang heuristic is to define variable
polarity as a phase with the maximum weight. The weight of a
variable depends on the number of clauses containing that variable
and their sizes. In \cite{chenPhase:22}, we tried to select a phase
of a variable, using the ACE (Approximation of the Combined
lookahead Evaluation) weight \cite{March:18,MoRsat:15}. This is a
dynamic policy. Its computation is time-consuming. Therefore, this
policy is not so successful, but can be applied to a part of SAT
instances in a way similar to portfolio methods. Glucose adopts a
phase selection policy based on the RSAT heuristic: it always
assigns a decision variable to false if that variable was never
visited, and the previous value otherwise. Such a phase selection
policy is simple, but in some cases, we found it is not so
efficient.

  The goal of this paper is to find a new phase selection heuristic
that improves the phase selection of such solvers as Glucose. If we
can select always correctly a phase, all satisfiable formulae will
be solved in a linear number of decisions. In theory, no perfect
phase selection heuristic exists unless P=NP. In practice, it is
possible to develop a phase selection heuristic that significantly
reduces the number of conflicts in some cases. To reach conflicts as
soon as possible, a dynamic weight seems to be suited for the phase
select of variables. A dynamic weight of a literal at a decision
level is defined as the sum of weights of its implied literals at
that decision level.Implied literals are assignments that were
forced by BCP. So the definition of a weight here is actually close
to that of the ACE weight \cite{March:18,MoRsat:15,chenPhase:22}.
Our new phase selection strategy is based on this dynamic weight. In
general, computing dynamically a weight is time-consuming. Hence, so
far no SAT solver applies successfully a dynamic phase selection.
However, since the implied literal of our strategy conforms to that
of the search process, the usual two watched-literals scheme can be
applied to the weight computation. Thus, our dynamic weight can be
computed efficiently, and apply to the phase selection of decision
variables. Empirical evidences show that our new phase selection
scheme can improve the performance of such state-of-the art solvers
as Glucose and Lingeling \cite{Precosat10:5}.

\section{A Dynamic Phase selection}
In modern conflict-driven SAT solvers, How to select the phase  of a
variable is an inseparable step that follows the decision variable
selection, because we must assign each decision variable to a value.
The simplest phase selection policy is that each decision variable
is always assigned to false, which is used as a default heuristic of
MiniSAT. No evidence shows that such a policy is always efficient.
Therefore, other policies are adopted in some solvers. For example,
PrecoSAT \cite{Precosat10:5} used Jeroslow-Wang heuristic. Here, we
present a new dynamic phase selection policy. Its dynamic weight is
based on a static weight. Let ${\cal F}$ define an input formula in
CNF (Conjunctive Normal Form) clauses. The static weight of a
literal $x$ on ${\cal F}$ is defined as

 \hskip 10mm $W(x,{\cal F})=\sum\limits_{c \in {\cal F}(x)} 5^{2-\mathrm {size}(c)}$

\noindent where ${\cal F}(x)$ is the set of clauses in which $x$
occurs, and $\mathrm {size}(c)$ is the size of clause $c$. This is
very similar to the definition of a weight in Jeroslow-Wang
heuristic \cite{JWheuris:2}. The main difference between two is that
the base is different. Our base is 5, while the base of
Jeroslow-Wang heuristic is 2. Selecting 5 here is based on the fact
that the March solver uses also base 5 \cite{March:18}. In this
paper, we define the dynamic weight of a literal $x$ as the sum of
the static weight of literals implied by it. This definition can be
formulated as follows.

 \hskip 10mm $DW(x,{\cal F},{\cal F}')=\sum\limits_{{x \wedge \cal F}'\vdash y} W(y,{\cal F})$

\noindent where ${\cal F}$ and ${\cal F}'$ are an input formula and
a formula at a search state, respectively. Usually, ${\cal F}$ is
constant, and ${\cal F}'$ varies with the search state. $x \wedge
{\cal F}'\vdash y$ means that using the fact that $x$ is true,
applying unit resolution on formula ${\cal F}'$ can derive an
implication $y$. That is, $y$ is an implied literal of $x$ under
${\cal F}'$. Computing implied literals is simple. This can be done
by a unit propagation, i.e. so-called BCP. Thus, without developing
a new routine, we can use directly a ready-made BCP of a SAT solvers
to compute implied literals. The dynamic strategy here is different
from that used in \cite{chenPhase:22}. The dynamic strategy in
\cite{chenPhase:22} needs such an additional data structure as a
full watched-literals scheme. The dynamic strategy here need not any
additional data structure, and can apply directly a two
watched-literals scheme. Therefore, our dynamic strategy is very
efficient. Once a variable is decided, the dynamic strategy elects
the branch with the highest dynamic weight $DW$. $W(y,{\cal F})$ can
be computed in advance. In addition, computing implied literals of a
literal is consistent with BCP. Therefore, we integrate the
computation of dynamic weights with the search procedure in
conflict-driven SAT solvers. Let $x$ be the decision variable. A
search on $x$, including the computation of dynamic weights, may be
described as follows.

\begin{flushleft}
\hskip 10mm {\bf search ($x$,$W,{\cal F}'$)} \\
\hskip 14mm $\langle Y_+$, Ret$\rangle \leftarrow$ BCP($x$, ${\cal F}'$)\\
\hskip 14mm if Ret=UNSAT then return UNSAT \\
\hskip 14mm backtrack to current level \\
\hskip 14mm $\langle Y_-$, Ret $\rangle \leftarrow$ BCP($\neg x$, ${\cal F}'$)\\
\hskip 14mm if Ret=UNSAT then return UNSAT \\
\hskip 14mm if $DW(W,Y_-) > DW(W,Y_+)$ then return $\neg x$ \\
\hskip 14mm backtrack to current level \\
\hskip 14mm BCP($x$, ${\cal F}'$) \\
\hskip 14mm return $x$\\
\end{flushleft}

\noindent In the above procedure, the parameter $W$ is used to store
the static weights of all the literals. ${\cal F}'$ is the current
formula, which can be maintained usually by a trail tack. $Y_+$ and
$Y_-$ are the set of literals implied by $x$ and $\neg x$,
respectively. In addition to fulfilling two tasks of the usual BCP:
compute the implied literals and determine whether it reaches a
conflict, BCP($x$, ${\cal F}'$) returns  the set $ Y_+$ of literals
implied by $x$. The usual search runs only one time, but our search
need to run at most three times. If the dynamic weight of $\neg x$
is larger than that of $x$, we run BCP just two times, since in such
a case, the last BCP is consistent with the search direction.
Clearly, in the worst case, the cost of our search is at most triple
the cost of the usual search if each BCP has the same cost.

 Always selecting the phase with the highest weight may lead to an early contradiction.
Thus, applying fully the above dynamic policy may be profitable in
solving unsatisfiable formulae, but not necessarily favours solving
satisfiable formulae. To trade off the performances on unsatisfiable
formulae against the performances on satisfiable formulae, we
combine the above dynamic policy and other phase selection policies.
In our SAT solvers, we divide the whole solving process into several
search periods. A search period refers to the search process between
two restarts. The notion of restarting is from the work of Gomes
\emph{et al} \cite{restart:1997}. Its meaning is that the solver
abandons its current partial assignment and starts over again.
Restarting is now considered as an essential component of modern
backtracking SAT solvers. In different search periods, we can use
different phase selection policies. However, in a search period, we
use only one phase selection policy. Furthermore, any phase
selection policy does not change the restart policy. Below we
present a few phase selection schemes, which will be used to improve
Glucose.

\begin{enumerate}

\item F+Save scheme: This scheme always
assigns a decision variable to false if no previous value was saved,
and the previous value otherwise. Its phase saving policy is to save
the values of visited variables at only the last decision level when
backtracking.

\item T+Save scheme: This scheme is the same as the previous scheme
except for the initial phase value. That is, it always assigns a
decision variable to true if no previous value was saved, and the
previous value otherwise.

\item F+All\_save scheme: This scheme is the same as the F+Save scheme
except for the phase saving policy. The phase saving policy of this
scheme is to save the values of visited variables at all the last
decision levels when backtracking. This is actually the phase select
policy of Glucose.

\item Odd-Even dynamic scheme: This is a hybrid scheme. It interchanges the
static policy  with the dynamic policy. During the odd numbered
search periods, it uses the above dynamic phase selection policy at
the odd numbered decision levels, and does the F+Save policy at the
even numbered decision levels. During the even numbered search
periods, it uses the above dynamic phase selection policy at the
even numbered decision levels, and does the F+Save policy at the odd
numbered decision levels. Its phase saving policy is the same as the
F+Save scheme.

\item Bit-encode scheme: This scheme lets the phase at each decision
level correspond to a bit value of the binary representation of an
integer. Assume that the binary representation of $n$  is

\hskip 8mm $n=b_k2^k +b_{k-1}2^{k-1}+\cdots+b_12+b_0$.\\
This scheme stipulates that during the $n$-th search period, the
phase of a variable at the $k$-th decision level is equal to $b_k$.
Usually, only the first 6 decision levels uses this scheme. And the
other levels uses such a policy as the Odd-Even dynamic scheme.

\end{enumerate}

To improve better the performance of such SAT solvers as Glucose, in
our SAT solver, we select the phase of a variable in the following
way.

\begin{description}

\item [(1)] For a large formula, say, its number of literals is greater than
1600000, within the first 1000000 conflicts, we use the F+All\_save
scheme. Once the number of fixed variables in a search period
exceeds $1\%$,  we switch to Odd-Even dynamic F+Save, T+Save and
scheme at the subsequent $(3k)$-th, $(3k+1)$-th and $(3k+2)$-th
($k=0,1,\ldots$) search period, respectively. After the 1000000
conflicts, we continue to use the F+All\_save scheme.

\item [(2)] For a small formula, i.e., its number of literals $\leq
1600000$, in general, we use Odd-Even dynamic scheme. However, if at
the 600000-th conflicts the number of fixed variables is still
smaller than 3, we switch to the combination of Bit-encode scheme
and the other schemes. In details, In such a case,  we use T+Save,
F+Save and Odd-Even dynamic scheme at the subsequent $(3k)$-th,
$(3k+1)$-th and $(3k+2)$-th ($k=0,1,\ldots$) search period, but does
the Bit-encode scheme at the first 6 decision levels except for the
Odd-Even dynamic scheme.

\item [(3)] When the number of conflicts reaches 5000000, for any formula, we use
interchangeably T+Save, F+Save and Odd-Even dynamic scheme. In
details, we use these schemes at the the subsequent $(3k)$-th,
$(3k+1)$-th and $(3k+2)$-th ($k=0,1,\ldots$) search period,
respectively.

\end{description}

 As the search process proceeds, the number of fixed variables
 increases generally. This will result in that the input formula ${\cal F}$
 can be simplified constantly during the solving process. Many SAT solvers make use of a
 simplifying process. The static weight based on the simplified
 formula should be different from that based on the original
 formula. Therefore, we re-compute the static weights every time the formula ${\cal F}$
 is simplified. However, if the refresh frequency of the formula ${\cal F}$
 is too high, we give up some computations on the static weights to
 save the solving cost. In details, we remove the re-computing of the static weights where the
 number of conflicts between two simplifications is less than
 200000. That is, only when the formula ${\cal F}$ is updated and the
 number of conflicts between two updates $> 200000 $, we update the static
 weights.

\section{Empirical evaluation}

\begin{table}
\caption{ Runtime (in seconds) required by Glucose 2.0 and improved
Glucose to solve some application problems.}
\begin{center}

\setlength\tabcolsep{4pt}
\begin{tabular}{l|c|c|c|c}
\hline  \hline
\multicolumn{1}{c|}{Instance} & \# var & \# clauses & Glucose 2.0 &
improved \\
&  &  &  & Glucose \\
\hline
slp-synthesis-aes-bottom14 & 22886 & 76038 & $>$9000 & 545.6\\
slp-synthesis-aes-top26  & 76943 & 245006 & 3794.3  & $>$9000 \\
slp-synthesis-aes-top28  & 88763 & 282870 & $>$9000 & 3263.5 \\
slp-synthesis-aes-top29  & 94998 & 302862 & $>$9000 & 4546.3 \\
minxorminand128 & 153834 & 459965 & $>$9000 &  8398.7 \\
gss-22-s100.cnf & 31616 & 95110 & $>$9000 & 7246.8 \\
AProVE07-01 & 7502 & 28770 & $>$9000 & 8349.1 \\
eq.atree.braun.12.unsat & 1694 & 5726 & $>$9000 & 6139.9 \\
comb1 & 5910 & 16804 & $>$9000 & 479.6 \\
rand\_net70-60-10 & 8400 & 25061 & $>$9000 & 820.0 \\
k2fix\_gr\_rcs\_w8 & 10056 & 271393 & 4134.0 & $>$9000 \\
vmpc\_35 & 1225 & 211785 & 6586.7 & $>$9000 \\
vmpc\_36 & 1296 & 230544 & $>$9000 & 1463.9 \\
 \hline
\end{tabular}
\end{center}
\end{table}

\begin{table}
\caption{ Performance of solvers on 300 application instances in SAT
2011}
\begin{center}

\setlength\tabcolsep{4pt}
\begin{tabular}{l|c|c}
\hline  \hline
\multicolumn{1}{c|}{Solver} & Instances Solved & Average time (in seconds) \\
 & & per solved instance \\
\hline
Glucose 2.0  &  214 & 1023.1 \\
Improved Glucose   & 221 & 998.1 \\
\hline
\end{tabular}
\end{center}
\end{table}

\begin{figure}
\centering
\includegraphics[height=7.2cm]{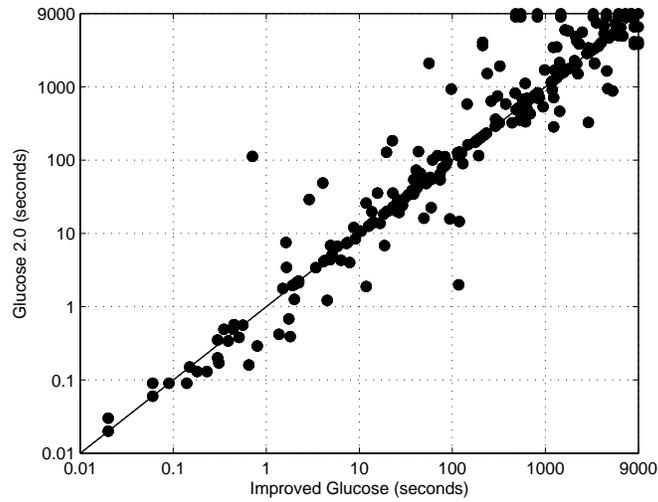}
\caption{Comparing the runtimes of Glucose 2.0 and improved Glucose
on application instances from SAT 2011.} \label{singleFig}
\end{figure}

\begin{figure}
\centering
\includegraphics[height=7.2cm]{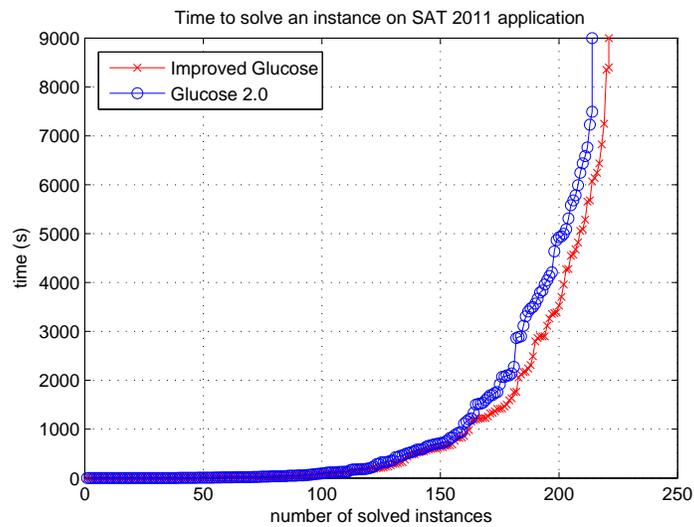}
\caption{The number of instances that Glucose 2.0 and improved
Glucose are able to solve in a given amount of time. The x-axis
denotes the number of solved instances, while the y-axis denotes
running time in seconds. } \label{singleFig2}
\end{figure}

We evaluated the new phase selection strategy, using the following
experimental platform: Intel Core 2 Quad Q6600 CPU with speed of
2.40GHz and 2GB memory. This is a 32-bit machine. From our empirical
results, this machine seems to be about half the speed of the
experimental platform used by SAT 2011 competition. Therefore, we
set the timeout for solving an instance to 9000 seconds, which is
almost double that of SAT 2011 competition. The instances used in
the experiment are from SAT 2011 competition. We use Glucose 2.0,
which won a Gold Medal for application category in the SAT
competition 2011, to test performance of the new phase selection
strategy. The main different between our improved Glucose and the
original Glucose is that the improved Glucose adopts the phase
selection scheme given in the previous section. The preprocessing of
the improved Glucose is the same as that of Glucose 2.0. Both
solvers use SatElite for preprocessing.

Table\,1 shows instances that solved by Glucose 2.0 and the improved
Glucose. Among the 13 instances, 3 instances were solved by only
Glucose 2.0, while the other 10 instances were solved by only the
improved Glucose. except for the 13 instances, on application
category in SAT 2011 competition, the numbers of instances solved by
the two solvers are the same. Table\,2 presents the number of solved
instances and the average running time per solved instance in
seconds. Glucose 2.0 and the improved Glucose solved 214 and 221 out
of 300 application instances, respectively. In terms of the average
running time, the improved Glucose was a little faster than Glucose
2.0.  On application instances, the performance of Glucose 2.0 here
is consistent with that in SAT 2011 competition, except for
slp-synthesis-aes-top29. This instance was not solved by Glucose 2.0
on our machine. This may be because the double precision of our
machine is different from that of SAT 2011 competition.

Figures 1 shows a log-log scatter plot comparing the running times
of Glucose 2.0 and the improved Glucose on application instances
from SAT 2011. Each point corresponds to a given instance. The
climax (9000,9000) means that the instances on that point were not
solved by any of two solvers.  As shown in Figures 1, many points
are centralised at the nearby diagonal. This is because on about 50
instances, both solvers use the same phase selection policy. Points
below the diagonal correspond to instances solved faster by Glucose
2.0.  It is easy to see that the number of instances solved faster
by Glucose 2. 0 is less than that solved faster by the improved
Glucose.
  Figure 2 shows a cactus plot related to the comparison of the two solvers.
Clearly, our phase selection strategy outperforms that of Glucose
2.0. In the cactus plot, our curve is always below Glucose 2.0. That
is, in a given amount of time, we solved more instances than Glucose
2.0.

\begin{table}
\caption{ Runtime (in seconds) required by Lingeling 587f and
improved Lingeling to solve some application problems.}
\begin{center}

\setlength\tabcolsep{4pt}
\begin{tabular}{l|c|c|c|c}
\hline  \hline
\multicolumn{1}{c|}{Instance} & \# var & \# clauses & Lingeling 587f
&
improved \\
&  &  &  & Lingeling \\
\hline
slp-synthesis-aes-bottom14 & 22886 & 76038 & $>$9000 & 4780.3\\
bobsmhdlc2-tseitin & 44692 & 129620 & 5335.5 & $>$9000\\
pdtvisns3p02-tseitin & 163622 & 488120 & 3808.7 & $>$9000 \\
countbits128-tseitin & 95810 & 287045 & 8993.4 & $>$9000 \\
minandmaxor128 & 249327 & 746444 & $>$9000 & 7099.1 \\
partial-10-13-s & 234673 & 1071339 & $>$9000 & 5284.0 \\
vmpc\_32 & 1024 & 161664 & $>$9000 & 2338.2 \\
vmpc\_34 & 1156 & 194072 & $>$9000 & 181.4 \\
vmpc\_35 & 1225 & 211785 & $>$9000 & 4548.0 \\
SAT\_dat.k85 & 181484 & 890298 & $>$9000 & 2056.6 \\
 \hline
\end{tabular}
\end{center}
\end{table}

\begin{table}
\caption{ Performance of solvers on 300 application instances in SAT
2011}
\begin{center}

\setlength\tabcolsep{4pt}
\begin{tabular}{l|c|c}
\hline  \hline
\multicolumn{1}{c|}{Solver} & Instances Solved & Average time (in seconds) \\
 & & per solved instance \\
\hline
Lingeling 587f  &  208 & 993.8 \\
Improved Lingeling   & 212 & 1080.5 \\
\hline
\end{tabular}
\end{center}
\end{table}

To improve Lingeling, we devise the following two phase selection
schemes.
\begin{enumerate}

\item full dynamic scheme: This scheme always
assigns a decision variable to a polarity with the highest dynamic
weight.

\item half dynamic scheme: This scheme always assigns a
decision variable to a polarity with the highest dynamic weight if
no previous value was saved, and the previous value otherwise.

\end{enumerate}

In the improved Lingeling, we select the phase of a variable in the
following way.

\begin{description}

\item [(1)] For a large formula, say, its number of literals is greater than
1500, in the first stage we use the half dynamic scheme. In the
second stage, we use the full dynamic scheme. From the third stage
to the end, we use the half dynamic scheme again. The maximal
numbers of conflicts at the first and second stage are limited to
300000 and 100000, respectively.

\item [(2)] For a small formula, i.e., its number of literals $\leq
1500$, in the first stage we use the full dynamic scheme. In the
second stage, we use the half dynamic scheme. From the third stage
to the end, we use the full dynamic scheme again. The maximal
numbers of conflicts at the first and second stage are limited to
10000 and 490000, respectively.

\end{description}

We conducted also experiments on Lingeling to justify the
effectiveness of the dynamic phase select policy.  Table\,3 shows
instances that solved by Lingeling 587f and the improved Lingeling.
Table\,4 presents the number of solved instances and the average
running time per solved instance in seconds. Compared with the
previous experimental results, it is easy to see that the
improvement on Lingeling is a little poorer than the improvement on
Glucose. The improved Lingeling solved only 4 instances more than
Lingeling 587f did. However, in terms of the average running time,
the improved Lingeling was a little slower than Lingeling 587f. This
maybe because the Jeroslow-Wang policy of Lingeling is close to our
dynamic phase selection policy.

\section {Conclusions and Future work}

  To improve the performance of conflict-driven SAT solvers, we have
developed a new dynamic phase selection policy. Unlike the ACE
dynamic weight used in \cite{chenPhase:22}, the new dynamic weight
is simple. And its computation cost is low. So, it is easy to embed
 the new phase selection policy in modern SAT solvers. Empirical results
 demonstrate that our new phase selection policy
can improve significantly the performance of solvers.

  Is a phase selection policy related to the other components such
as the restart policy, the learnt clause management policy, etc?
This is an open problem that is worth studying.

Another important is how to improve the new phase selection policy
and combine it and the existing phase selection policy. As a future
research subject, we will study it further.

Whether in theory or in practice, we believe that the phase
selection policies known so far are not certainly the best. However,
does there exist the best phase selection policy? If exist, how do
we find out it? This is a very valuable research topic.

\bibliographystyle{splncs}
\bibliography{DynamicsPhase12}

\end{document}